**ORIGINAL PAPER**

# An Experimental High-Throughput to High-Fidelity Study Towards Discovering Al–Cr Containing Corrosion-Resistant Compositionally Complex Alloys

Debashish Sur[1,2] · Emily F. Holcombe[3,4] · William H. Blades[5] · Elaf A. Anber[3] · Daniel L. Foley[3] · Brian L. DeCost[6] · Jing Liu[7] · Jason Hattrick-Simpers[6,8] · Karl Sieradzki[5] · Howie Joress[6] · John R. Scully[1,2] · Mitra L. Taheri[3]

## Abstract
Compositionally complex alloys hold the promise of simultaneously attaining superior combinations of properties, such as corrosion resistance, light-weighting, and strength. Achieving this goal is a challenge due in part to a large number of possible compositions and structures in the vast alloy design space. High-throughput methods offer a path forward, but a strong connection between the synthesis of an alloy of a given composition and structure with its properties has not been fully realized to date. Here, we present the rapid identification of corrosion-resistant alloys based on combinations of Al and Cr in a base Al–Co–Cr–Fe–Ni alloy. Previously unstudied alloy stoichiometries were identified using a combination of high-throughput experimental screening coupled with key metallurgical and electrochemical corrosion tests, identifying alloys with excellent passivation behavior. The alloy native oxide performance and its self-healing attributes were probed using rapid tests in deaerated 0.1-mol/L $H_2SO_4$. Importantly, a correlation was found between the electrochemical impedance modulus of the exposure-modified air-formed film and self-healing rate of the CCAs. Multi-element extended x-ray absorption fine structure analyses connected more ordered type chemical short-range order in the Ni–Al 1st nearest-neighbor shell to poorer corrosion resistance. This report underscores the utility of high-throughput exploration of compositionally complex alloys for the identification and rapid screening of a vast stoichiometric space.

**Graphical Abstract**

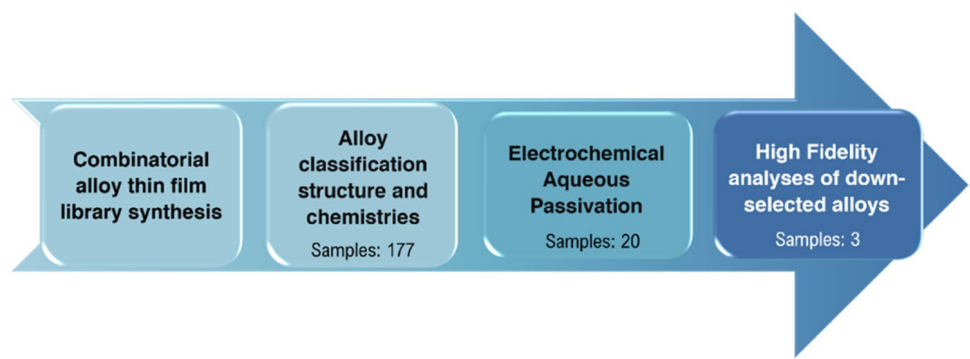

---

Debashish Sur and Emily F. Holcombe have contributed equally.

John R. Scully and Mitra L. Taheri have contributed equally.

Extended author information available on the last page of the article

## Introduction

With the advent of compositionally complex alloys (CCAs, also commonly referred to as high-entropy alloys or multi-principal element alloys), there has been strong motivation to investigate and explore the large multicomponent alloy phase design space [1–3]. Given the wide range of available compositions and potentially unique structures, their properties can be uniquely tailored. This incentivizes the exploration and optimization for various properties, such as promising mechanical behavior [1, 4–6], aqueous corrosion resistance [7–9], wear resistance [10], high-temperature oxidation resistance [11], and radiation tolerance [12], including sustainability of metals by alloy design [13, 14]. Despite several successful efforts to predict popular structure property relationships in high-entropy alloys, such as yield strength [15], no pathway exists to computationally screen alloys for good aqueous passivation performance. As Miracle points out [16], there are over 592 billion CCAs with bases of 3–6 principal elements (incrementally varying both element and composition). Such an enormous range of potential alloys requires a high-throughput (HTp) approach and targeted down-selection process to screen alloy compositions with unique structures and properties [17, 18].

Novel systems with potential properties of interest can be identified computationally and explored experimentally. Computational and machine learning (ML) approaches have been leveraged with success in predicting properties unrelated to corrosion [19–24]. Studies of ML descriptor selection and optimization for corrosion prediction have also been undertaken and enabled corrosion rate and pitting potential predictions [25, 26]. However, ML systems often base their predictions on inputs that are seemingly disconnected from physically defensible explanations of passivation behavior, such as interstitial concentration having an outsized impact on corrosion resistance or $H^+$ reduction potentials. For example, it has been posited that low $H^+$ reduction potentials are detrimental toward corrosion resistance [26]. Alternatively, other descriptors that would likely be more informative for ML-based modeling, such as the maximum driving force for solid metal oxide formation energy compared to solubilized metal cation formation energy [27], can be generated for an individual alloy. It is time consuming to compute these using density functional theory in sufficient quantity to enable training of ML models. However, there exists a critical need for corrosion data, both to train ML models as well as to validate model predictions.

High-throughput methods have gained popularity recently due to their ability to screen hundreds of compositions in alloy systems in a relatively short period of time [17, 18]. Rapid experimental testing methods and machine learning have been successful in screening increasingly complex systems, such as metallic glasses and CCAs [28–30]. While a variety of experimental techniques have been employed for alloy development depending on the system and property of interest, there still exists a significant need for the capability and methodology to conduct HTp experiments and then down select to specific alloys of interest based on utilization of rapidly acquired electrochemical metrics for aqueous passivation. The combination of computational and experimental HTp assessment of stable, single-phase regions enable the exploration of strong compositional contributions to structural properties, corrosion inhibiting coatings, metal dissolution rates, and passivation performance [31–35].

In the search for lightweight corrosion-resistant CCAs to date, research has primarily focused on selective additions of known passivating elements to base alloys, such as Cr, Al, and Si [7, 36–40]. However, the effects of these additions can be quite convoluted. For instance, the addition of Al to complex CoCrFeMnNi (Cantor) type bulk alloys has been shown to decrease the density but adversely affect the corrosion resistance of the alloys, typically leading to pitting-type behavior [41, 42]. This is not surprising considering the strong role Al plays in the formation of secondary phases, including the Ni–Al-rich B2 phase in an FCC matrix, thus leading to partitioning of Cr and Al [43]. However, for single-phase Fe-based alloys, addition of Al in solid solution with Cr has been found to be beneficial in chloride-free environments [44]. Moreover, hybrid Molecular Dynamics/Monte Carlo simulations revealed that addition of Al to Cantor-type alloys alters Cr–Cr bond lengths and can introduce Cr–Cr chemical short-range ordering (CSRO) [45]. Such preferred Cr–Cr localized bonding has been shown in other studies to influence the primary stage of passive film formation on binary Fe–Cr alloys in 0.1-mol/L $H_2SO_4$ [46]. This phenomenon has shown potential for tuning the performance of corrosion-resistant CCAs but is unproven [47]. As local CSRO tends to increase with addition of Al in these systems [45], an ability to screen compositions within a tight parameter space will be critical to understanding corrosion behavior going forward. Moreover, individual Cr and Al additions have been shown to favor low passive current densities in binary Fe–Cr [48] and Fe–Al [49] alloys in sulfuric acid. For Cr and Al containing single-phase CCAs, CSRO can be one of several physical aspects that affect alloy passivation and thus it is interesting to develop protocols to rapidly consider such possibilities.

Here, we demonstrate an experimental pathway toward down selecting corrosion-resistant Al–Cr–Fe–Co–Ni single-phase CCAs selected as a subset of a larger population of alloys over a range in compositions and structure. Our focus is on conducting high-fidelity analyses of chosen alloys

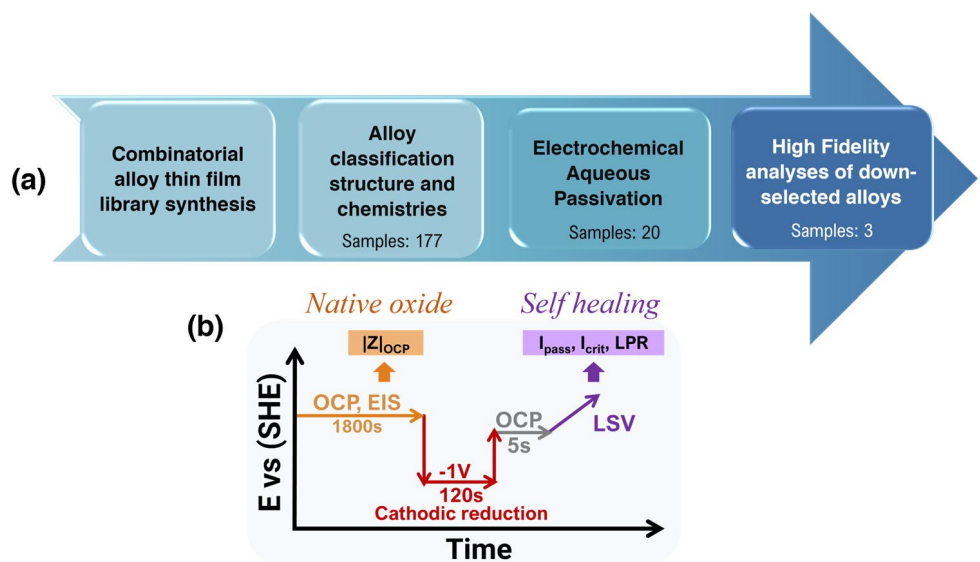

**Fig. 1** **a** A high-throughput to high-fidelity process flowchart a was followed in this work starting with thin-film library synthesis and continuing towards detailed analysis of down-selected alloys. The number inside the boxes represents the number of alloys studied in that stage. **b** The electrochemical sequence that was followed for high-throughput aqueous passivation performance studies

to explore the relationships between passivation, CSRO, microstructure, and the composition of passive films. The down-selection was informed by a HTp screening approach achieved using the combinatorial thin-film library synthesis technique in concert with rapid microstructural and aqueous corrosion evaluations. Further, we introduce a simplified approach to perform the difficult extended x-ray absorption fine structure (EXAFS) analyses of a five-component CCA and calculate Warren–Cowley (WC) parameters of interest. Combinatorial thin films CCAs, despite their unique energetic landscape as compared to their bulk counterparts, provide us with a pathway for rapid assessment of aqueous passivation. Through this approach, we underscore the importance of HTp exploration as it offers a path forward to rapidly identify promising corrosion-resistant alloys.

## Experimental Methods

A process flow outlining the pathway from high-throughput to high-fidelity methods is shown in Fig. 1a. The alloys identified by x-ray diffraction (XRD) were further investigated on diced wafers isolating single alloys using a rapid HTp approach following the protocol described herein. Alloy compositions were disqualified for these aqueous passivation experiments if they exhibited a clear presence of potentially detrimental secondary phases by XRD. Ultimately, the three compositions chosen for high-fidelity measurements were the alloys with the best two and the single worst passivation performance. These samples were analyzed further with high-fidelity analysis, including x-ray photoelectron spectroscopy (XPS), scanning transmission electron microscopy (STEM), and EXAFS measurements, providing a lens through which each alloy's passivation performance was correlated to phase, structure, and local chemistry.

## Combinatorial Thin-Film Deposition, Composition, and Microstructure

Combinatorial thin films were synthesized by magnetron co-sputtering from three sources onto a 76 mm (3 in.) Si wafer with a thermally grown $SiO_2$ layer. The geometry of the sputter guns (Mag Keeper, Kurt J. Lesker) was such that the centers of depositions for the three sources were off-center from the wafer and the deposition rate for each source varied as a function of position across the wafer, forming a continuous variation in composition. In this case, alloys were synthesized by magnetron co-sputtering (Mag Keeper, Kurt J. Lesker)[1] from a set of metallic alloy targets of $Al_{0.70}Ni_{0.15}Fe_{0.15}$, $Co_{0.70}Ni_{0.15}Fe_{0.15}$, and $Cr_{0.70}Ni_{0.15}Fe_{0.15}$, each with purities of better than 99.99%, with a power of 4.82 W/cm$^2$, 3.01 W/cm$^2$, and 3.01 W/cm$^2$, respectively. The deposition time was 2 h and the wafer was annealed at 400 °C for 2 h inside the chamber. Compositions within the film were determined by x-ray fluorescence (XRF) using a Bruker Tornado, analyzed by Crossroads XRS-MTFFP software, and verified by wavelength and energy-dispersive spectroscopy, the details of which can be found in the Supplementary.

---

[1] Certain commercial equipment, instruments, or materials are identified in this paper in order to specify the experimental procedure adequately. Such identification is not intended to imply recommendation or endorsement by the authors and their respective institutions, nor is it intended to imply that the materials or equipment identified are necessarily the best available for the purpose.

For high-throughput structural analysis, XRD patterns were taken using a Bruker D8 diffractometer using a Dectris Eiger area detector and Cu-Kα radiation of wavelength 0.154 nm. Scans were radially integrated to produce 1D diffraction patterns. Diffraction spots were measured with a 0.5-mm incident beam in a $\theta$–$2\theta$ geometry. A total of 177 diffraction spots each linked to a XRF composition and spot on the wafer [50, 51]. The classification of points followed a similar peer-adjudicated process as described elsewhere [52]. A "BCC*" designation represented alloys that were nominally single phase with no discernible B2 super lattice and/or FCC reflections, indicating they existed in at most negligible proportions. A "*Multiphase*" spot had a prominent (100) superlattice reflection at 31° indicating a B2 structure that is an ordered BCC phase [53] or a bifurcated peak around 45° representing FCC (111) and BCC (110) reflections.

Imaging of the wafer surface and composition verification of the down-selected samples was performed using a Thermo Fisher Scientific Helios UC G4 scanning electron microscope (SEM). STEM samples were prepared using a Thermo Scientific Helios G4 UC Focused Ion Dual-Beam SEM and images were obtained using a Thermo Fisher Scientific TF30 STEM equipped with a Fischione High-Angle Annular Dark Field detector. Chemistry, phase, and texture analysis were obtained using STEM-EDS (energy dispersion spectroscopy, EDAX Elite detector) and through precession electron diffraction via the Nanomegas ASTAR platform, yielding phase, grain size, and texture/orientation mapping.

## Aqueous Passivation

Positions on the wafer grid coordinate system indicating the category of BCC* were diced into chips of dimensions (4.5 mm × 4.5 mm) with a diamond blade dicing tool in a clean room. A flat three-electrode electrochemical cell contained a mercury/mercurous sulfate (0.640 V vs. SHE) reference electrode, a Pt mesh counter electrode, and the CCA thin film was used as the working electrode (projected exposure area of 0.06 cm$^2$) defined by a Viton O-ring. All electrode potentials in this work are reported vs. SHE. The measurements were conducted using a Gamry Reference 600 + potentiostat. The electrolyte was 0.1-mol/L $H_2SO_4$ solution prepared with ultrapure water (Millipore, 18.5 MΩ·cm) and the solution was deaerated by bubbling $N_2$ throughout the duration of the experiment. The sequence shown in Fig. 1b followed Fig. 4 for the rapid electrochemical survey experiments consisting of linear sweep voltammetry (LSV) and potentiostatic electrochemical impedance spectroscopy (EIS). The same experimental protocols were repeated on arc melted and homogenized bulk 316L stainless steel ($Fe_{0.70}Cr_{0.20}Ni_{0.09}Mo_{0.01}$) and equiatomic CoCrFeMnNi alloy coupons after polishing and cleaning them to a 1-µm

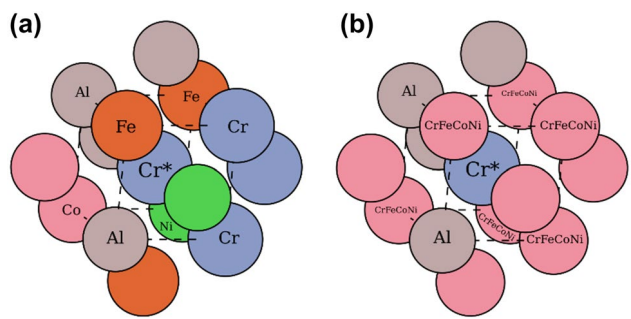

**Fig. 2** A schematic diagram of Cr local environment used to parameterize the EXAFS models. Dashed lines indicate the BCC unit cell with the absorbing atom Cr* at the body-centered site. **a** Without simplifying assumptions, the atomic cluster used for EXAFS analysis must contain all five species in the first (annotated with chemical symbols) and second neighbor shells. **b** A pseudo binary structural model with composite species for all transition metal scatterers M = {Cr, Fe, Co, Ni}

surface finish. The corresponding high-fidelity single-step 10-ks passivation and EIS measurements were performed at 0.6 V at adjacent positions on the thin film as well as for the bulk alloy coupons after cathodically reducing the air-formed oxide at − 1 V for 120 s.

## Coordination Analysis with EXAFS

Down-selected alloys were also measured using EXAFS to characterize the local coordination environments at each composition. EXAFS measurements were performed at the Beamline for Materials Measurements (BMM-6BM) at the National Synchrotron Light Source II (NSLS-II) using a micro-focused beam and a Si (111) monochromator. EXAFS signal was collected in fluorescence mode and normalized to the incident intensity with a $N_2$ ion chamber. The K-edge for each element in the films was measured with the exception of Al, which falls out of range of the monochromator. We used a simplified structural model illustrated in Fig. 2 to measure the nearest-neighbor Cr–Al and Ni–Al Warren–Cowley (WC) parameters $\alpha_{Cr-Al}$ and $\alpha_{Ni-Al}$ [54, 55]. A full description of our EXAFS analysis methodology can be found in the Supplementary section of this work. Our simplified model treats the four transition metal scattering elements in the first coordination shell (Fig. 2a) as a single aggregate transition metal scattering species M (Fig. 2b) to reduce the number of fitting parameters to something tractable. Due to this simplification, we are unable to directly measure WC parameters between pairs of transition metals (*e.g.*, $\alpha_{Cr-Cr}$ or $\alpha_{Cr-Ni}$). The WC CSRO parameter is a convenient tool for CSRO analysis because a value of $\alpha_{ab} = 0$ indicates no statistical neighbor preference between species A and B, $\alpha_{ab} > 0$ indicates fewer $A - B$ neighbors than in a random configuration (clustered CSRO), and $\alpha_{ab} < 0$ indicates more $A - B$ neighbors (ordered CSRO) than random [56].

## Chemical Composition of Passive Film

Chemical compositions and valence states of passive films grown on down-selected alloys were assessed using XPS after the sequence of cathodic reduction and a single-step passivation for 10 ks at 0.6 V. The samples were transferred from the electrochemical cell to the XPS chamber in a nitrogen-filled glove bag within a few minutes after the passive film growth experiments. XPS core shell spectra were collected using Al Kα x-rays (1,468.7 eV) at a 26-eV pass energy, 45° take off angle, and 100 × 100-μm spot size using a PHI VersaProbe III system. For Fe and Co, the Auger peaks overlap with their core-level $2p_{3/2}$ ranges which required acquisition of high-resolution spectra focusing on the Al 2p, Cr $2p_{3/2}$, Fe $2p_{1/2}$, Co $2p_{1/2}$, Ni $2p_{3/2}$, and O 1s core-level energies [57]. The small overlap of Cr 3s in Al 2p core levels was accounted for by assuming its amplitude proportional to the Cr $2p_{3/2}$ signal, relative to their sensitivity factors [58]. The details of the spectral fitting procedure can be found elsewhere [59–61].

## Results

### High-Throughput XRD Classification

The combinatorial thin film was sampled at 177 evenly spaced positions, identified by the coordinate system with a unique composition and structure as described in Fig. 3. The aluminum, chromium, and cobalt varied between 20 and 40 at.%, 15–35 at.%, and 15–35 at.%, respectively. The Ni and Fe compositions were roughly fixed at 15 at.%. Each alloy was analyzed for structure and composition. The resulting categorized compositions were mapped on to a pseudo-ternary phase diagram. Several single-phase alloys, categorized as BCC*, were found to be in the region of the composition diagram with Al and Cr contents above approximately 20 at.% but below 30 at.%. The remaining positions exhibited a multiphase structure. BCC* alloys were selected for subsequent HTp corrosion testing.

### High-Throughput Electrochemical Survey

Rapid aqueous electrochemical passivation experiments were performed following the sequence shown in Figs. 1b and 4. The sequence covers two important aspects of corrosion-resistant alloy design, their native oxide performance, and capability to self-heal by forming a passivating layer that limits corrosion in a harsh electrolyte. For exposure

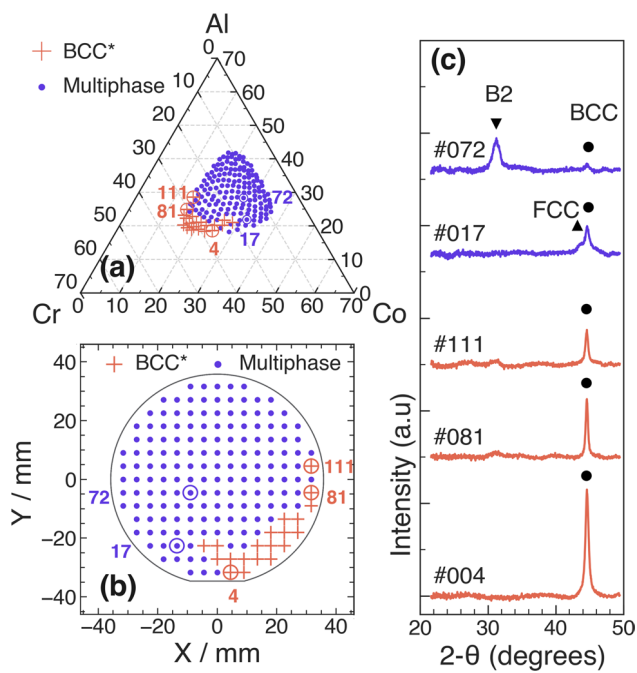

**Fig. 3** **a** Compositional and **b** Spatial representation of HTp microstructural phase classification based on thin-film XRD and XRF of $Al_{0.7-x-y}Co_xCr_yFe_{0.15}Ni_{0.15}$ library after annealing at 400 °C for 2 h. **c** HTp XRD patterns of alloy spots #4, #17, #72, #81, and #111 showing the different phases acquired using a Cu Kα X-ray source

modified native oxide performance, electrochemical impedance modulus at 10 mHz ($|Z|_{OCP}$) was determined from an EIS frequency sweep after 30 min at OCP [62]. A trend of increasing $|Z|_{OCP}$ was observed across all the BCC* alloy regions with the increasing alloy Al + Cr content as shown in Fig. 4a. From their bare alloy anodic LSV data, linear polarization resistance (LPR) [62], passive current density ($i_{pass}$), and critical current density ($i_{crit}$) were extracted that are defined metrics of corrosion and self-healing nature. Further details of LPR calculations can be found in supplementary material. The $i_{pass}$ was taken as the lowest current density in the passivation region [63, 64]. An increasing $|Z|_{OCP}$ correlated with a lower $i_{crit}$, $i_{pass}$ and higher LPR values as shown in Fig. 4b. Hence, $|Z|_{OCP}$ was regarded as a rapidly obtained corrosion metric (i.e., requiring a few hundred seconds test time) for rapidly forecasting alloy passivity behavior.

Alloys were ranked based on their respective $|Z|_{OCP}$ values and the best and worst performing alloys were down-selected among the BCC* category for further examination. Figure 4c and d shows the EIS at OCP and LSV behaviors of the two best (#81, #111) and the lowest performing (#4) alloy compositions, respectively. These behaviors are compared to bulk 316L stainless steel and a CoCrFeMnNi (Cantor) alloy. These electrochemical parameters are tabulated in Table 1. #81 and #111 have a comparable $|Z|_{OCP}$ with a slightly greater $i_{pass}$ and $i_{crit}$ in comparison to bulk 316L

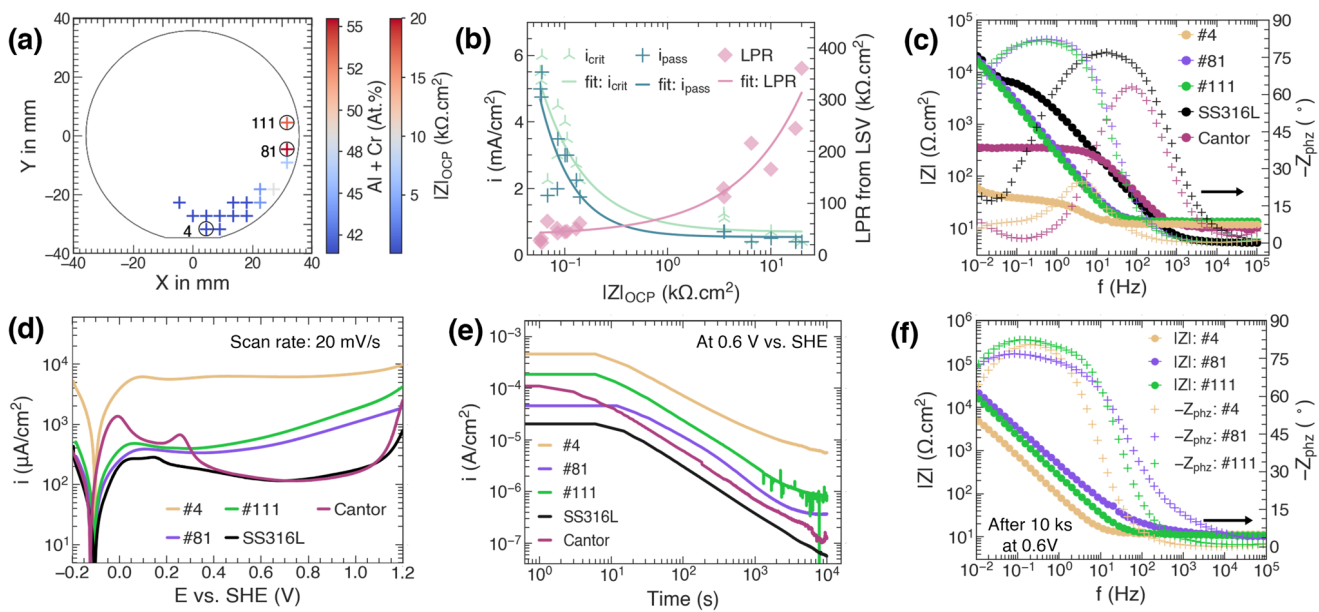

**Fig. 4** **a** Spatial variation of $|Z|_{OCP}$ as native oxide performance after exposure for 30 min in deaerated 0.1-mol/L $H_2SO_4$, and (Al+Cr) at.% across the BCC* alloys on the wafer grid. Encircled spots are the down-selected spots #4, #81, and #111. **b** Electrochemical parameter correlations of $i_{pass}$, $i_{crit}$, and LPR obtained from LSV performed after cathodic reduction at −1 V for 120 s with $|Z|_{OCP}$ of the native oxide films. **c** Bode magnitude and phase plots of EIS at OCP of exposure-modified native oxide after 30 min, and **d** upward LSV after cathodic reduction of down-selected alloys #4, #81, #111, 316L stainless steel and the CoCrFeMnNi (Cantor alloy) in deaerated 0.1-mol/L $H_2SO_4$. **e** Current density decay of down-selected alloys and 316L stainless steel and Cantor during potentiostatic hold at 0.6 V vs. SHE during passive film growth for 10 ks after cathodic reduction step, and **f** full-spectrum EIS Bode magnitude and phase angle plots at 0.6 V vs. SHE of #4, #81, and #111 alloys after the 10 ks potentiostatic hold, all in deaerated 0.1-mol/L $H_2SO_4$

**Table 1** Three down-selected composition spots of $Al_{0.7-x-y}Co_xCr_yFe_{0.15}Ni_{0.15}$ library labeled as point numbers and bulk control alloys with their respective alloy compositions, structure, and measured electrochemical behavior metrics in deaerated 0.1-mol/L $H_2SO_4$ at stages 'EIS' and 'LSV' are indicated in Fig. 4a

| Alloy | Stoichiometry | Structure (XRD) | $\|Z\|_{OCP}$ at 10 mHz (kΩ cm²) | LPR (kΩ cm²) | $i_{pass}$ (mA/cm²) | $i_{crit}$ (mA/cm²) |
|---|---|---|---|---|---|---|
| #111 | $Al_{0.28}Co_{0.15}Cr_{0.27}Fe_{0.15}Ni_{0.15}$ | BCC*[a] | 17.5 | 295 | 0.4 | 0.5 |
| #81 | $Al_{0.25}Co_{0.15}Cr_{0.30}Fe_{0.15}Ni_{0.15}$ | BCC* | 20 | 364 | 0.4 | 0.4 |
| #4 | $Al_{0.18}Co_{0.25}Cr_{0.27}Fe_{0.15}Ni_{0.15}$ | BCC* | 0.060 | 21 | 5.5 | 6 |
| Cantor | $Co_{0.2}Cr_{0.2}Fe_{0.2}Mn_{0.2}Ni_{0.2}$ | FCC | 0.375 | 209.6 | 0.175 | 1.5 |
| 316L | $Cr_{0.20}Fe_{0.70}Mo_{0.01}Ni_{0.09}$ | FCC | 20 | 681.6 | 0.150 | 0.25 |

[a]The "BCC*" designation would represent an alloy having BCC as the majority phase with second phases like B2, FCC being negligible in phase fraction due to their extremely low-intensity or unclear XRD peaks

stainless steel. When compared to the bulk Cantor alloy, they both outperform it based upon $|Z|_{OCP}$ and $i_{crit}$ but have a greater $i_{pass}$. This $i_{pass}$ does not account for a surface area factor in sputter-deposited alloys which ranged from 2 to 5x. This would make results similar if implemented.

## High-Fidelity Passivation Performance

To confirm the trend shown in the HTp corrosion study, higher-fidelity passivation studies were performed. Figure 4e shows the potentiostatic current decay in the passive range at 0.6 V after cathodic reduction for 10 ks. It indicates lower current densities after 10 ks and the reduction of the total anodic charge required to repassivate alloys #81 and #111 compared to alloy #4. Clearly, alloy #4 showed inferior performance from the standpoint of the passive current density magnitude and the time for the current density to decay below 10 μA/cm² (passive current density ≤ 10 μA/cm² is an excellent indicator of a stable passive film that regulates the oxidation rate). This repassivation time required is over 1000 s for #4 compared to just a few hundred seconds for alloys #111 and #81. This behavior retains

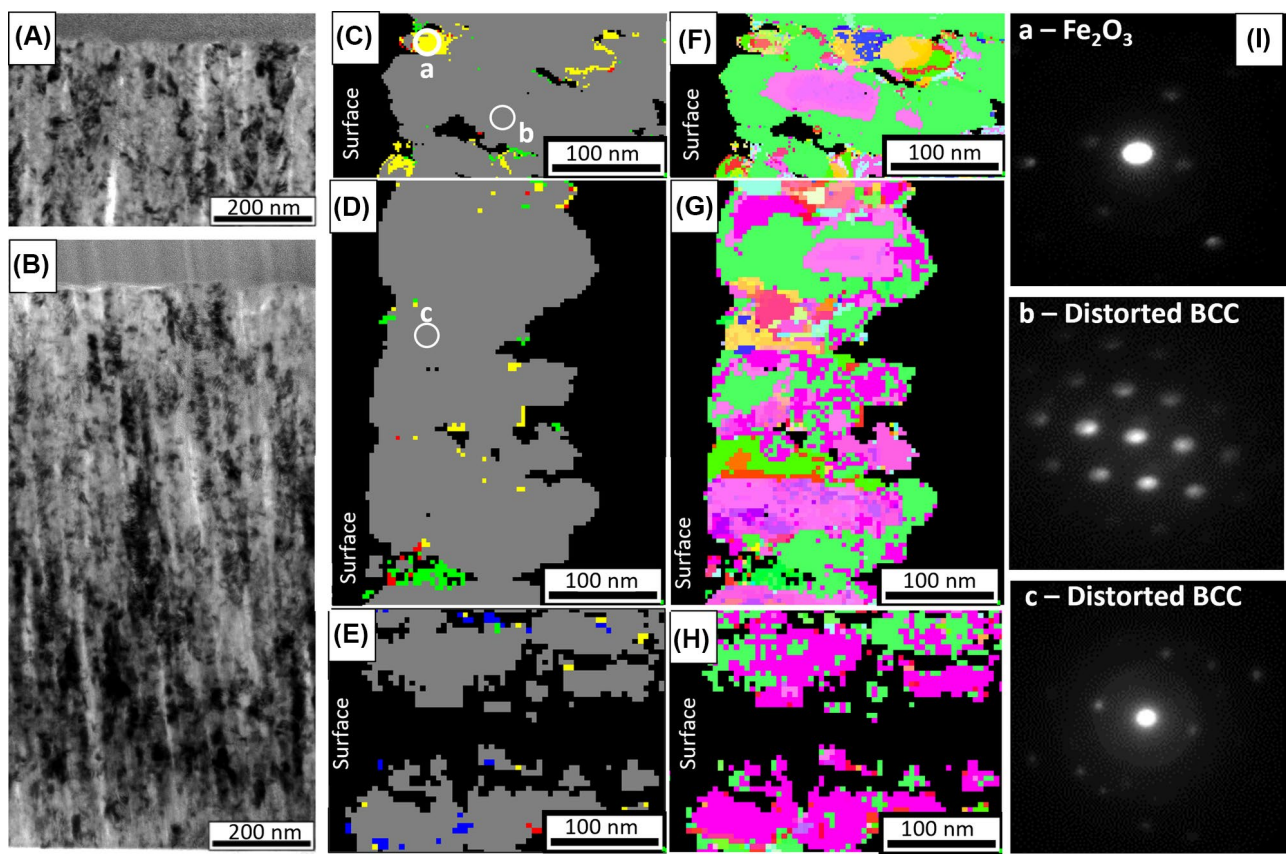

**Fig. 5** **A**, **B** Brightfield image of the #81 and #111 surface and through-thickness, respectively, using transmission electron microscope. **C–E** Precession Electron Diffraction (PED) results for #81, #111, and #4, respectively. The phase maps of #81 and #111 are based on indexing of reference structures for aluminum oxide (red), chromium oxide (green), iron (III) oxide (yellow), and a distorted BCC reference (gray). The phase map for #4 also includes indexing to an FCC reference (blue). **F–H** Inverse Pole Figure (IPF) maps of #81, #111, and #4, respectively, with colors associated with indexed orientations relative to the [001] direction. **I** Representative diffraction patterns that were indexed to produce the phase map in **C**, **D**. **I.a** Diffraction pattern that indexed to aluminum oxide, and **I.b** and **I.c** are representative distorted BCC diffraction patterns (Color figure online)

the trend established with respect to $|Z|_{OCP}$ as observed in repassivation of these alloys (Fig. 4b). The comparison to 316L stainless steel and Cantor alloy does not contain true surface area correction. The best CCAs here may approach 316L stainless steel but likely retain a slightly higher current density. A subsequent full-spectrum EIS at 0.6 V after 10-ks potentiostatic hold is shown in Fig. 4f of the downselected alloys. Comparing the impedance at 10 mHz, the trend remains consistent with HTp findings given the lower impedance of the solution formed passive film of #4 compared to #81 and #111.

## High-Fidelity Characterization of Morphology and Structure

Preliminary micrography using SEM revealed that the thin-film alloy surface resembled a "mud-cracked-like" surface where columns of solid material are separated by networked surface cracks as shown in Figs. S2 and S3. Quantitative analysis of the top surface void area fraction samples #81 and #36 was 13.2% and 15.1%, respectively, indicating relative consistency between alloys in the film. The #36 alloy is within the single-phase region and is approximately halfway between #81 and #4 on the wafer grid and is within the single-phase region. The through-thickness TEM brightfield images of samples #81 and #111 as shown in Fig. 5a and b demonstrate a consistent morphology through the thickness of the film with no distinguishable surface layer post-exposure. Both #81 and #111 had similar grain sizes, but #4 has a smaller average grain size of 12.7 nm vs 14.7 nm. The average grain size for each of the three compositions can be found in Supplementary Table S1. The through-thickness grain structure as shown in Fig. 6 varies with visibly elongated grains in #111 versus #4. Structural analysis in TEM using precession electron diffraction (PED) confirmed a BCC matrix phase with possible oxides as shown in Figs. 5c and 6b. These oxides were not factored into the electrochemical results as they were observed away from the surface.

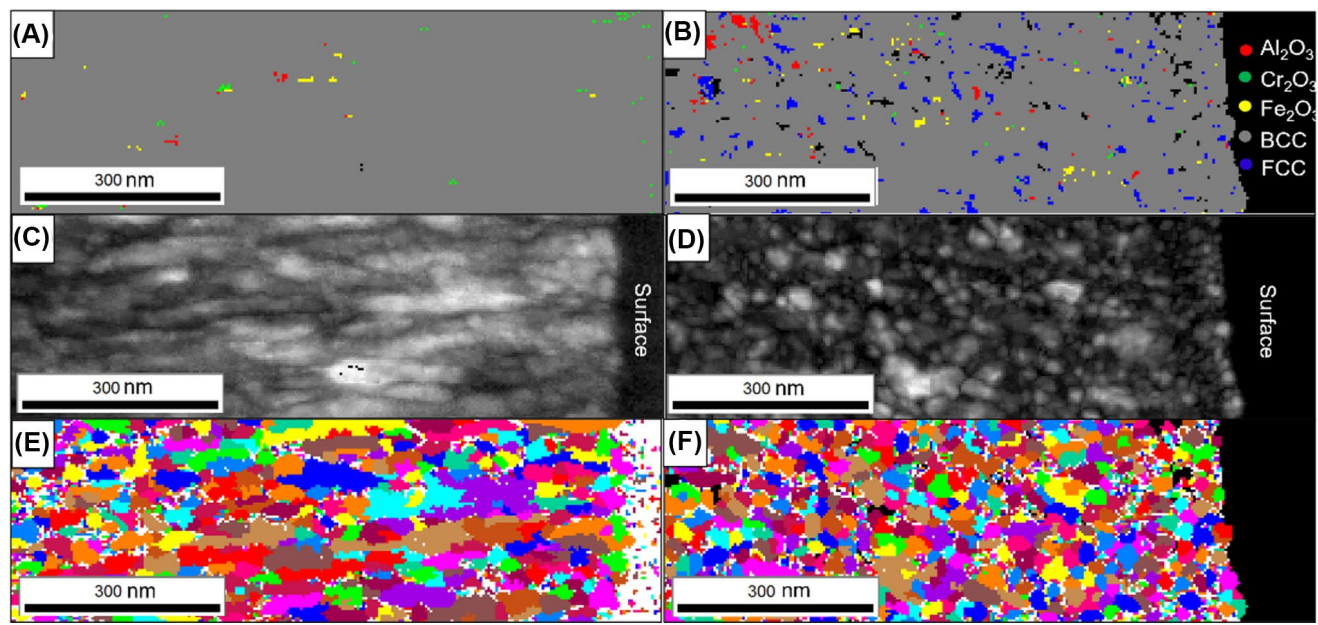

**Fig. 6** Scanning Transmission Electron Microscopy (STEM) phase maps using precession electron diffraction (PED) demonstrating oxide dispersion in the through-thickness of samples #111 (**A, C, & E**) and #4 (**B, D, & F**). Both contained a distribution of aluminum (red), chromium (green), and iron (yellow) oxides (**A & B**). The image quality (**C & D**) and orientation mapping show grain size from precession indexing are shown (**E & F**). FCC results are presented in Supplementary Fig. S11. **C & D** Image quality maps of indexed points used for PED. Signal quality is reduced at apparent grain boundaries and near the substrate. **E & F** Unique grain color maps produced following a grain dilation procedure for #111 and #4, respectively (Color figure online)

Despite regions of the film indexing to oxide structures, oxygen was found to be largely homogeneous through STEM-EDS (electron-dispersive spectroscopy) as shown in Supplementary Fig. S6. FCC was included in indexing but was not expected in #81 or #111 based on the x-ray diffraction results and thermodynamics (refer Supplementary sections). When included as a potential phase, some diffraction patterns in #81 and #111 will index to FCC structure. The observed diffraction patterns contain varying angles between reflections that challenge indexing and suggest distortion of the BCC matrix. Example diffraction patterns indexed as BCC vs. FCC are provided in the Supplementary Fig. S7. Effectively, the matrix phase for #81, #111 and #4 should be considered as distorted BCC and in #4 the through-thickness map in Fig. 6b suggests the FCC phase is approximately 4% by volume. This level is consistent with secondary phase content that can go unnoticed in XRD. While there is some uncertainty in the quantitative presence of oxides and FCC phase, the results confirm a primarily BCC matrix with no significant oxide layer at the exposed surface, matrix lattice distortion, and little evidence of secondary phases.

## High-Fidelity Characterization of Elemental Clustering by EXAFS

Figure 7 shows the Fourier-transformed EXAFS spectra (weighted by $k^2$) for the Cr, Fe, Co, and Ni K-edge on samples #4, #81, and #111 alloys plotted in real space. The EXAFS data were reduced and plotted for each edge along with the theoretical contribution of Al, Cr, and Co scattering species (Fe and Ni are not shown as they are virtually identical to Co) for the first and second nearest-neighbor shells. Full-edge Fourier-filtered spectra can be found in Supplementary information. This qualitative visualization uses equal bond length, MSRD (0.005 Å$^2$), and CN variables for all scattering paths. For these BCC structures, the scattering contributions of the first and second shells overlap in real space, so both needed to be considered during the fitting analysis as shown in Fig. 8. The scattering contributions of Al, Cr, and Co to the EXAFS signal of each absorbing species M* = (Cr, Fe, Co, and Ni) are represented as M*–Al, M*–Cr, and M*–Co for the first and second shell. For Cr, the measured EXAFS signal in Fig. 7a was nearly identical for Sample #81 and #111 and much lower in magnitude for #4. The quantitative analysis is summarized in Table 2. The Cr–Al WC parameter was found to be positive and close to zero for both 1st and 2nd NN shells, indicating only a slight suppression of Cr–Al pairs. The Ni–Al WC parameters for both shells were negative, indicating a clustering tendency between Ni and Al. For the Fe edge, shown in Fig. 7b, all three measured samples were virtually identical. Conversely, for the Co (Fig. 7c) and Ni (Fig. 7d), there were strong differences in the EXAFS results between the three samples.

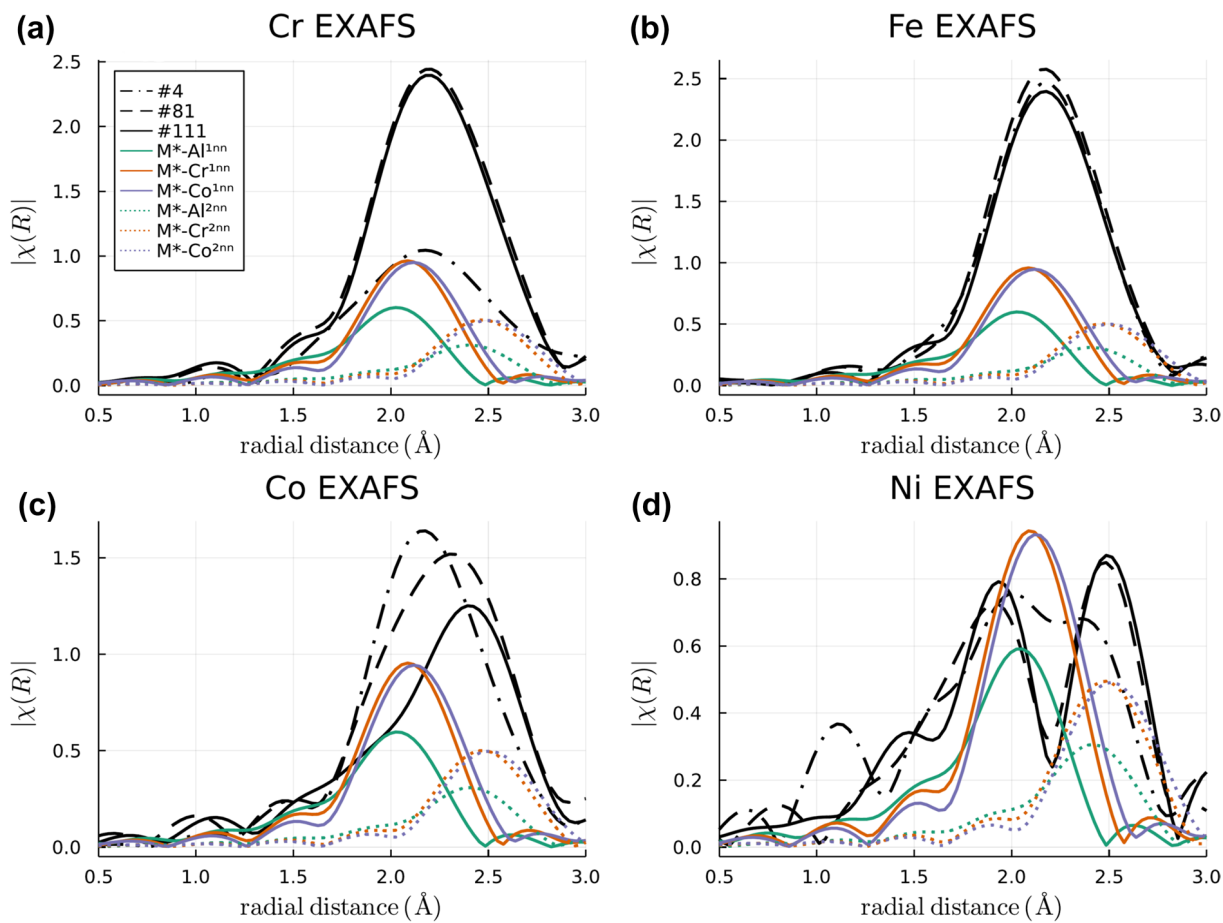

**Fig. 7** Reduced and Fourier-transformed (R-space) EXAFS spectra for the **a** Cr, **b** Fe, **c** Co, and **d** Ni K edges for alloys #4, #81, and #111 in black. M*–Al, M*–Cr, and M*–Co represent the scattering contributions of Al, Cr, and Co from M* = (Cr, Fe, Co, and Ni) as the central absorbing species for the first (solid lines) and second (dotted lines) shell

We qualitatively evaluate the trends for these data in the Discussion section.

## High-Fidelity Characterization of Passive Films by XPS

Figure 9 shows the high-resolution XPS passive film cation fractions of the alloys #111, #81, and #4, associated with the core–shell spectral fits of the selected passivators Cr and Al and that of O of alloy #81 of the passive films potentiostatically grown at 0.6 V vs. SHE for 10 ks. Very little Ni, Co, or Fe was observed in the oxides measured on #111 and #81. A slightly greater concentration of Ni(II), Fe(II), and Co(II) are detected in oxides on alloy #4. A small contribution of $NiCr_2O_4$ (Ni–Cr spinel) [58, 60] was found to be consistent with a portion of Cr and Ni spectra, while maintaining stoichiometric ratio of normalized intensities. Cr and Al cations occupy at least 60 at.% of the cations in the passive films. Cr was enriched as much 2× relative to the bulk atom % of Cr in alloys #81 and #111, but not in #4. #81 and #111 also displayed the highest Cr + Al cation surface fractions. Unlike Cr, Al did not show an extensive enrichment. Alloy #4 showed the least enrichment of passivating Cr(III) and Al(III) cations in their surface passive films which correlated well with the bulk composition of the alloy, i.e., (Al + Cr) at. %.

## Discussion

### Thermodynamic and Microstructural Analyses

One of the major aims concerning thin-film CCAs is to synthesize alloys with a large library of compositions that are equivalent to cast bulk alloys. For our HTp analysis, the first considerations were the phases and compositions of each spot on the wafer. A calculated phase diagram (CALPHAD)-based prediction of stable phases is presented in the Supplementary section, which suggests multiple phases are present after annealing ($T = 400\ °C$) for all

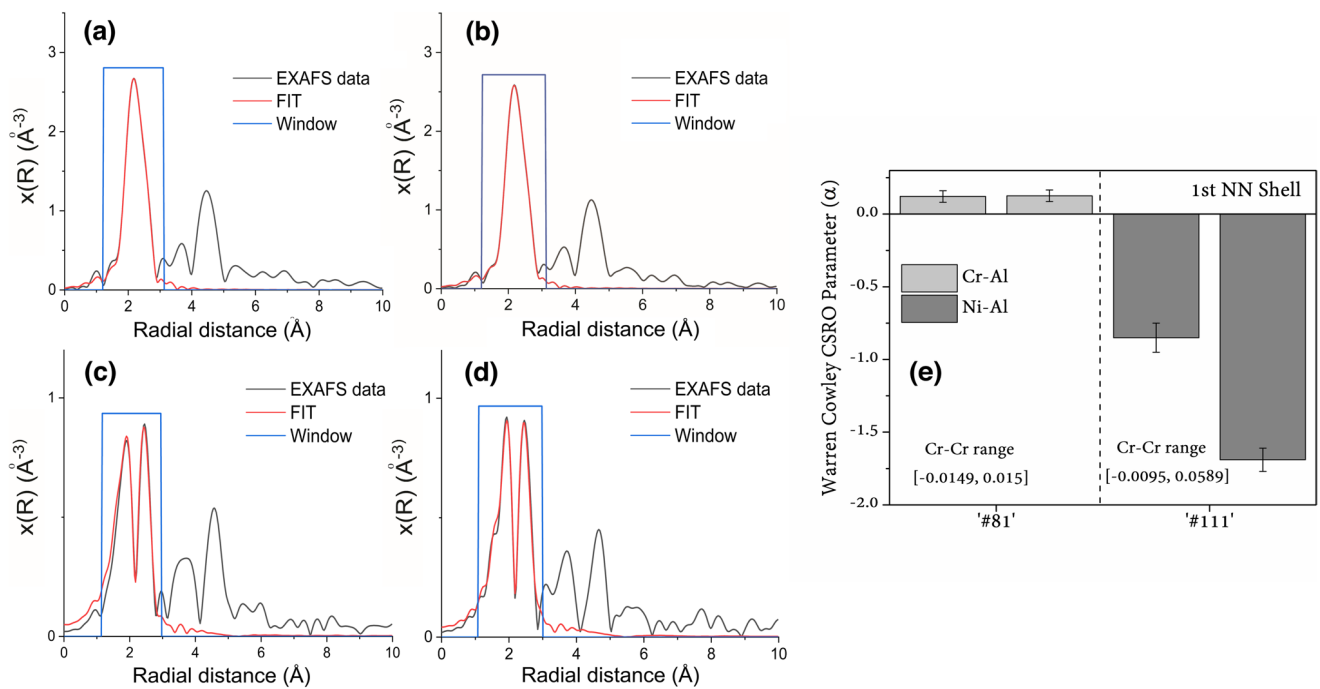

**Fig. 8** Fourier transform (r-space) Cr–K edge EXAFS spectra of **a** #81 and **b** #111 and Ni–K edge EXAFS spectra of **c** #81 and **d** #111, respectively. These spectra were used to determine the average coordination numbers (CNs). **e** Warren–Cowley (W–C) parameters of Cr–Al and Ni–Al interactions in 1st NN shell for alloys #81 and #111. Negative W–C parameter indicates attractive interactions between A and B atoms, while positive denotes repulsive interactions between A and B atoms

**Table 2** Coordination numbers and W–C parameter of samples #81 and #111, measured from the Cr edge using EXAFS

| #Alloy | Nearest neighbor | $\sigma^2_{Cr-M}$ (Å$^2$) | $CN_{Cr-M}$ | $\sigma^2_{Cr-Al}$ (Å$^2$) | $CN_{Cr-Al}$ | $\sigma^2_{Ni-Al}$ (Å$^2$) | $CN_{Ni-Al}$ |
|---|---|---|---|---|---|---|---|
| 81 | 1NN | 0.005 ± 0.003 | 6.25 ± 0.151 | 0.000 ± 0.004 | 1.749 ± 0.151 | 0.00024 ± 0.002 | 3.70 ± 0.2 |
| 111 |  | 0.004 ± 0.002 | 5.989 ± 0.166 | 0.000 ± 0.003 | 2.010 ± 0.166 | 0.021 ± 0.005 | 6.164 ± 0.2 |
| 81 | 2NN | 0.005 ± 0.003 | 4.505 ± 0.16 | 0.000 ± 0.004 | 1.49 ± 0.16 | – | – |
| 111 |  | 0.004 ± 0.002 | 4.248 ± 0.398 | 0.000 ± 0.003 | 1.751 ± 0.398 | – | – |

Uncertainties correspond to one standard deviation three down-selected alloys. Thermodynamically, the BCC/B2 phases are predicted for #81 and #111 and in addition, Fe/Co–Cr type sigma phase is predicted for #4. Interestingly, this contrasts the observed structure of our films, as a small peak of B2 was found in #111, while no such peak including that of sigma was observed in #81 and #4 (Fig. 3 and Supplementary Fig. S9). Indeed, it is unclear whether the precipitation and growth of the B2 is predictive of bulk behavior, as CCA thin films can have non-equilibrium amorphous or metastable microstructures [65]. Relative phase stability may offer insight into thermodynamic predictions, especially in the case of metastable films. It has been shown by OQMD [66] that increasing aluminum content stabilizes NiAl structures and destabilizes the FCC HEA phase [43]. The trend has been shown to hold true experimentally for the kinetically stabilized thin films with the stabilization of BCC/B2 over the FCC HEA phase in the high-aluminum compositions, including #4 where FCC is thermodynamically stable. Additionally, within the two thermodynamically BCC-only compositions, B2 was observed in #111 with 4 at% higher Al content, but not in #81 [46].

The classification of XRD peaks can be subjective regarding the influence of the background and noise on small deviations that could represent peaks. The peer-reviewed classification process adopted for screening alloys via HTp scanning XRD identified points with low confidence in the clear presence of any second phase peak. This was resolved by defining a common category BCC* as shown in Fig. 3, which encompasses all spots without the certain presence of any second phase while keeping in mind the kinetic

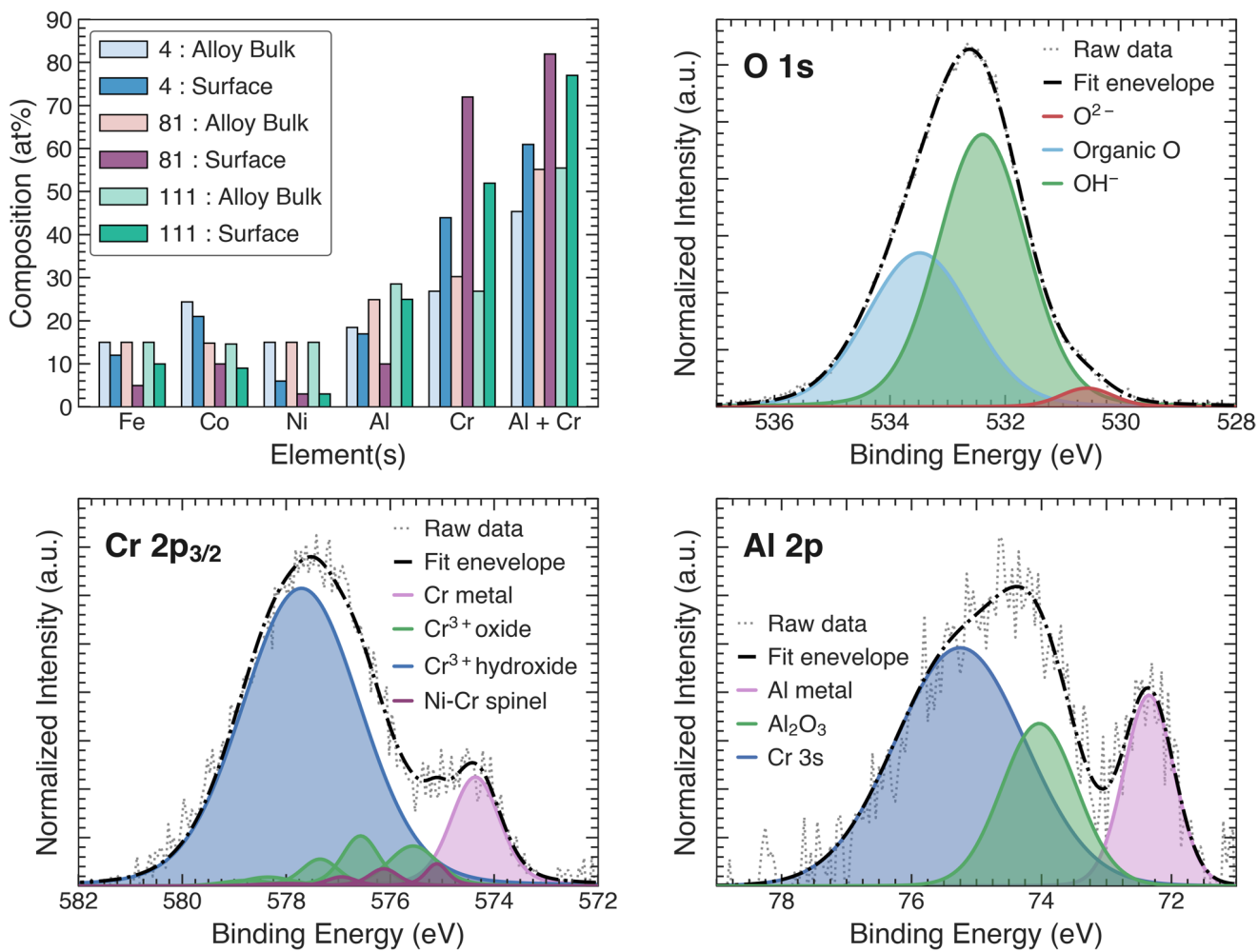

**Fig. 9** X-ray photoelectron spectroscopy (XPS) based passive film cation compositions of the down-selected alloys #4, #81, and #111, obtained from their core shell spectral deconvolution and peak fitting. An example of such is shown for the alloy #81 for its O 1s, Cr $2p_{3/2}$, and Al 2p core–shell XPS spectra. The passive film for these alloys was potentiostatically grown at 0.6 V vs. SHE for 10 ks in deaerated 0.1-mol/L $H_2SO_4$

stabilization of microstructures in thin films [30, 65], and the possible limited resolution of detecting small phase fractions with the chosen XRD method. For a high-throughput process, the finding of BCC* phases and their electrochemical behavior show enough confidence to support further high-fidelity studies into the performance of high-aluminum BCC HEAs and the role of B2 content in these metastable alloy phases. To be explicitly clear, evidence of secondary phases in the XRD could not be sufficiently quantitatively confirmed in this analysis, and the TEM structure quantification could be challenged by the lattice distortion [67]. The texture and low intensity of the B2 peak observed in the XRD results complicate the accurate estimation of B2 content. Subsequently, further work is needed to improve film quality (voids, recrystallization, texture, reduce, or normalize distortion) to likewise improve the quantification of phases and CSRO in the high-fidelity analysis.

In the present work, we grew kinetically stabilized BCC/B2 alloys as predicted by literature and show that this phase stabilization in sputtered films can be leveraged to explore the connection between chemical short-range order (CSRO) and passivation. A study by Kube et al. on the prediction of phase formation in HEAs found that BCC/B2 was kinetically stabilized in compositions where FCC was also a thermodynamically stable phase [30]. Their phase stability thresholds were based on the fitting of elemental contributions to phase formation from their combinatorial sputtering data, named the solid solution selection index (SSSI), and a simplified average considering elements as binary BCC or FCC stabilizers, called the FCC-BCC-Index (FBI). The authors found that a combination of FBI and difference in atomic size was a viable predictor of phase stability. Figure 10 shows the FBI and atomic size difference plot for our combinatorial library system as feedback on using the new

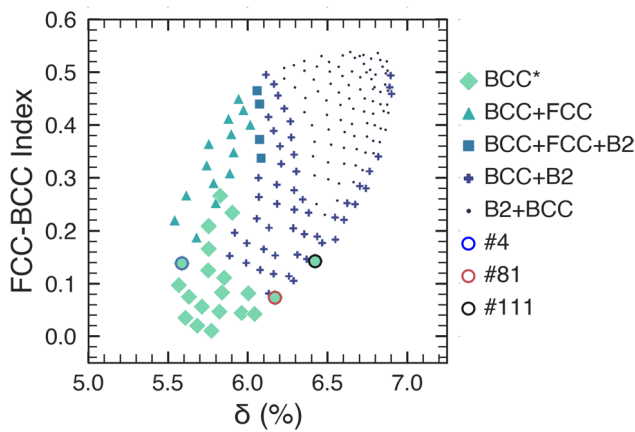

**Fig. 10** Calculated FCC-BCC-Index vs atomic size differences (δ) for all the 177 alloy compositions and classified from XRD data [30]. The three circled data points represent the three down-selected alloy compositions. The new categories include "BCC + B2", "B2 + BCC", "BCC + FCC," and "BCC + FCC + B2," where the phase listed first is considered the majority phase

parameter. The majority phase was determined by a simple comparison of the intensities of constituent reflections, BCC (110), B2 (200), and FCC (111). These observations are made on the lab XRD relative intensity data alone and should be confirmed with high-fidelity characterization or Rietveld analysis using idealized reference structures. Our B2 region is in alignment with the B2 region of the original work. The region of BCC + B2 extends further in FBI than the original work predicted, and the BCC + FCC region is well defined where there is significant overlap in the original work. The phase results for the three alloys studied agree with the SSSI phase predictor with #81 and #111 correctly predicted as BCC* phases with SSSI lower than –0.22 and #4 predicted as having some heightened probability of containing the FCC phase with an SSSI of − 0.15. These agreements demonstrate the applicability of the optimized descriptors for an off-stoichiometric sputtered CCA system [30]. As combinatorial material libraries evolve and begin to include consideration of chemical long-range order, such as B2 and CSRO, it may be possible to predict not only the single-phase structure but also CSRO. Further exploration of predictive descriptors that include chemical long- and short-range order, may offer an explanation for the difference in B2 expression between the alloy systems.

## High-Throughput to Higher-Fidelity Aqueous Passivation

A HTp corrosion testing sequence (shown in Fig. 1b) was explored here yielding more than four corrosion parameters for each BCC* alloy within an about 35-min test period on each alloy spot in 0.1-mol/L $H_2SO_4$. Deaerated 0.1-mol/L $H_2SO_4$ was the chosen solution environment to minimize the contributions of Fe, Co, and Ni toward alloy passivation while recognizing that high-purity Cr and Al could kinetically passivate [68], (Fig. S4). In an automated system, 24 samples could easily be tested in a 12-h period and will be explored in future [69]. A HTp equivalent circuit fitting was not possible for the EIS data due to lack of a reliable software package but can be possible in the future following a recently reported work [70]. EIS, as shown in Fig. 4c, was chosen to assess the sulfuric acid exposure-modified native oxide quickly after a 30-min exposure at OCP and is a sufficient period for the approximately 3-nm-thick oxides to either dissolve or grow when exposed to high acidity. Hence, the tests evaluated the short-term oxide stability and protectiveness by investigating whether it grew and repaired atomic sites where acid dissolution occurred spontaneously at a rate exceeding its dissolution rate.

The environment clearly should be extended to $Cl^-$ environments in future and that has been done but is not reported herein. More significantly, our HTp pathway highlights that a 35-min test is required to conduct similar corrosion screening tests. Additionally, the EIS-derived $|Z|_{OCP}$ of the exposure-modified native oxide (Fig. 4b) is shown to serve as a non-destructive preliminary screening parameter. This is because EIS at intermediate frequency is reflective of oxide thickness, porosity, and semi-conducting qualities reflecting ionic defects, and charge-compensating electrons or holes [71]. At low frequency, EIS is also reflective of the passive current density. By following the impedance measurements during single-step passivation, several additional parameters of value can be readily determined. After a cathodic reduction of the native oxide during the linear voltage sweep (Fig. 4d) or the potentiostatic hold (Fig. 4e, f), the primary passivation active–passive transition is driven by the selective dissolution of non-passivating d-block elements (e.g., Fe, Co, and Ni) and preferential passivation with Cr and Al [47]. When enough monomers of Cr–O–Cr or Al–O–Al bonds create a high surface coverage, the voltage gradient applied is distributed across the interfaces and the oxide thickens, producing a current density decay. This process may be affected by CSRO in the substrate alloy [46, 72]. The Al and Cr accumulated in the oxides while Fe and Ni could be incorporated as a result of solute captured or dissolved in solution, which is consistent with the passive film oxide compositions obtained herein from XPS. Thus, Al may play two roles: these are (a) affecting CSRO and Cr–Cr bond length and (b) secondary passivator to the Cr(III) primary passivator.

Passivation metrics involving potential-driven selective dissolution can be affected by the presence of voids or 'mud crack'-type features as shown in Supplementary Figs. S2 and S3. The first effect to consider is geometric effects on electrolyte concentration and mass transport. Concentrated sulfuric acid, unlike in the case of crevice corrosion in neutral chloride solution [73], would remain equally concentrated in acid in voids and likely not be further acidified by metal ion hydrolysis. Anodic passive current from voids would be similar to exposed surfaces because rates are regulated by the oxide, not mass transport in solution. However, the wetted area that the passive current comes from must be taken into account. Quantitatively, a similar increase in true exposed surface area compared to the projected area was observed from SEM image analysis across the region of interest of the wafer grid, details are provided in supplementary material. Further, the grain sizes obtained via TEM are not different within a significant order of magnitude. This suggests that the electrochemical behavior of the thin-film alloys in this region can be compared independent of their surface morphologies. LSV results show distinct electrochemical differences in tandem with $|Z|_{OCP}$ results, which showed very small differences between combi spot #81 and the bulk homogenized 316L stainless steel under similar conditions. The passive current densities of the alloys might be affected as discussed above. Such results demand the need to unravel these differences in a more detailed comparative study focused on the bulk vs thin-film passivation behavior of CCAs. However, relative changes in electrochemical behavior between spots on the combi-wafer are still of value when compared to one another and still offer significant insight into the electrochemical passivation of these alloy films.

## Chemical Composition of Passive Films: XPS

XPS was used to characterize the solution formed passive film after it was potentiostatically held at +0.6 V for 10 ks. A high concentration of Al and Cr cations in the passive film was measured by XPS indicating a good passivating alloy [48, 74, 75]. Unlike Fe, Co, and Ni that actively dissolve, oxides of Cr and Al can be observed in 0.1-mol/L $H_2SO_4$ environment. $Al_2O_3$ is thermodynamically unstable below a pH of 4 [76] but is spontaneously formed via kinetic mechanisms at lower potentials than pure Cr [77] (refer to Supplementary Fig. S4). Although Al oxide can be porous, it has the potential to form Cr oxide during Cr(III) injection as $Al_2O_3$ and $Cr_2O_3$ which may have complete mutual molecular solubility during the primary passivation stage, the $Cr_2O_3$–$Al_2O_3$ system likely co-exists in a single solid solution oxide [78, 79]. There is a tendency, however, for $Cr_2O_3$ and $Cr(OH)_3$ to dominate the passive film based on subtle differences in protection capability in this environment. However, varying Cr and Al cationic compositions could develop over 10 ks which could impact oxide and passive film homogeneity. A detailed study on the interplay between Al and Cr on primary and long-term passivation is needed. Further, as shown in Table 1 and Fig. 4f, the $i_{pass}$ and $|Z|$ at +0.6 V both show a direct correlation with the total fraction of Al(III)+Cr(III) cations in the passive film of #81 and #111 compared to #4 shown in Fig. 9. This correlation could be extended to all rapidly tested compositions and Al+Cr content as shown in Fig. 4a. Although Co content is higher in #4 compared to #81 and #111, its role here is not clear as it has been not been shown to affect passive film performance [61]. Overall, XPS cation fractions confirmed the electrochemical quality of these film spots as determined by $|Z|_{OCP}$ and LSV and offer a useful route toward additional non-destructive high-fidelity measurements post-passivation. In keeping with such analysis, high Al(III) and Cr(III) in alloys #111 and #81 forecast greater protectiveness seen in other parameters.

## Chemical Short-Range Order: EXAFS

Due to the difficulty of quantitative EXAFS analysis particularly in solid solutions with several species of similar atomic number [80], a qualitative comparison of the first- and second-shell EXAFS spectra of the alloys was considered in addition to a quantitative fit to the Cr and Ni edges as described above, shown in Figs. 7 and 8. There exists a complex nonlinear relationship between local structure and the resulting EXAFS signal. Furthermore, the data for alloy #4 must be interpreted in the context of the XRD evidence of the presence of two phases; the EXAFS will be a weighted average over the absorbing atoms in each phase. As expected, because Al has a much lower atomic number than the other elements, the Al atoms have significantly weaker scattering for all edges and therefore a set of shells containing large amounts of Al will have a less intense peak in the EXAFS.

For the Fe edge (Fig. 7b), all three measured samples had similar EXAFS spectra and therefore the trend in the Fe–Al Warren–Cowley (WC) [55] parameters is expected to be dominated by the Al composition. The measurements of the Co edge (Fig. 7c) suggest varying amounts of Al among the three samples, consistent with their average Al composition. This also implies that most of the additional Al in #111 compared to #81 is accommodated in the local environment of the Co atoms. As the Al concentration increases, the data suggest that long-range ordering of the Al around the Co atoms increases with a high amount in the first NN shell and a lesser amount in the 2nd NN shell, consistent with the XRD measurements (Fig. S9). The Ni edge was more difficult to analyze due to interference of the varying scattering paths, leading to the appearance of two peaks. Roughly, it showed a variation consistent with changing Al content in the 1st and 2nd NN shells; however, in this case, the trend was the opposite of the expected Al content based on the average composition of the samples. Given that #81 and #111 have similar Cr edge EXAFS spectra, it was surmised that the local atomic structure around Cr atoms will be alike in those two samples. #4 had a significantly less intense peak in the EXAFS suggesting that there is a large amount of Al in the first two NN shells around Cr, despite that alloy containing much less Al. The WC parameter results from the quantitative fitting of the Ni and Cr edges for #81 and #111 shown in Fig. 8e that corroborates the predicted presence of CSRO in Fe–Co–Ni–Cr–Al-type CCAs with Al addition [45]. Both #81 and #111 showed a positive Cr–Al WC parameter of + 0.12 and a very large negative Ni–Al WC parameter, − 0.85 and − 1.69, respectively, indicating strong Al–Ni clustering.

The relatively Al-rich environment around #4 suggests limited Cr–Cr clustering relative to #81 and #111. Clustering of like passivating elements (e.g., Cr–Cr) has been predicted to benefit primary passivation processes [47], while the clustering of unlike elements (e.g., Cr–Fe) has been shown to delay passivation [46]. Also, CSRO effects were shown to be less significant at Cr concentrations above the classical threshold, for a bcc Fe–Cr system. [46] Yet for a five-component CCA, such CSRO effects may inherently prevail and contribute to the observed poorer performance of #4 in comparison to #81 and #111 (Fig. 4). Further, Ni–Al clustering could detrimentally impact the passivation behavior in a number of ways [7] including locally interfering the configurational benefit of Cr–Al interactions leading to Cr–Cr pairs. #111 showed a more negative WC for Ni–Al and that would explain its relatively poorer performance compared to #81. Given the amount of Al in the alloys, Ni–Al intermetallic phases such as B2 were expected to be present [43, 81], but no observable secondary phases were identified. Nevertheless, the EXAFS supported the notion that Ni–Al order-type CSRO was present, which in turn suggests such atomic configuration of Ni–Al can alter the percolation landscape of all atoms, including Cr–Cr [47] that not only maintains a single-phase alloy but also contributes to corrosion protection. Clearly, this nanostructure effect requires significant further inquiry but demonstrates the variety of factors affecting corrosion properties.

Here, we note some of the limitations of the quantitative EXAFS analysis. In addition to the assumption of equal bond lengths and MSRD parameters for all transition metals, the choice of Cr as the signature of the composite transition metal species does not perfectly represent the other three transition metal scatterers, Fe, Co, and Ni. This potentially biases the quantitative estimates of the structural parameters EXAFS fit. Improving the robustness of the EXAFS analysis to definitively distinguish the Cr–Cr ordering from Cr -{Fe, Co, Ni} will require development of robust multi-edge EXAFS analysis for disordered solid solutions, which we defer to future work.

## Conclusion

A CCA library was synthesized using magnetron sputtering and was characterized using high-throughput structural analysis, composition analysis, and corrosion assessments sensitive enough to detect the quality of their native oxide film and active–passive transition behavior (self-healing). The workflow resulted in a down-selection from 177 unique compositions to just two compositions of interest, while comparison was drawn to one of the inferior alloy compositions of the library. The electrochemical impedance modulus of acid exposure-modified native air oxide at OCP obtained via EIS was found to serve as a viable screening parameter. This provided a strategy/protocol for rapid interrogation of alloys for facile indication of desirable passivation qualities in sulfuric acid. Therefore, a high-throughput pathway toward the discovery of good passivation performance has been demonstrated. Modification to this approach will likely emerge especially when testing for pitting in chloride containing electrolytes.

The combinatorial wafer synthesis process successfully produced and identified a good passivating single-phase BCC structure with > 20 at.% Al combined with Cr and low cobalt (< 15 at.%) that may have similar resistance to sulfuric acid corrosion as 316L stainless steel. The passivation was traceable to high enrichments of Cr(III) and Al(III) in the passive film, their high theoretical mutual solubility, as well as minimization of Fe(II), Ni(II), and Co(II). Metallurgical factors, such as microstructure, were also of concern. While the most obvious considerations were multiple phases and alloying element partitioning, these were substantially similar herein in high and low performing alloys; however, ordering effects remain significant.

The quantitative EXAFS fits indicate a negative and positive CSRO of Ni–Al and Cr–Al, respectively. The alloy with more Ni–Al chemical short-range ordering showed a relatively poor passivation performance, supporting a previous work showing the effect of ordered type CSRO in Fe–Cr binary alloys [46]. Although the deposition process can produce films with complex topography, the ability of this process to produce homogeneous single-phase structures of materials with different compositions shows promise for future investigations. For example, the kinetic stabilization of single-phase BCC and avoidance of elemental segregation issues found in casting techniques provide an ideal sample for measuring CSRO. This could enable leveraging the CSRO of disordered thin-film alloys to create new or expand existing phase stability predictors while using descriptors, such as FBI and SSSI. These predictors along with good passivation performance criteria based on acquired electrochemical metrics, establish the opportunity to have an all-encompassing model of the role of CSRO in designing corrosion-resistant CCAs. In this screening study, Cr(III) is a passivator and Al(III) had roles as both a secondary passivator and as an element in solid solution which helps establish Cr–Al and Cr–Cr bond length.

**Supplementary Information** The online version contains supplementary material available at https://doi.org/10.1007/s44210-023-00020-0.

**Acknowledgements** The authors gratefully acknowledge partial funding from the Office of Naval Research (ONR) through the Multidisciplinary University Research Initiative (MURI) program (Award #: N00014-20-1-2368) with program manager Dr. Dave Shifler. D.S. was partially supported by the DMSE-UVA Olsen Graduate Fellowship for the duration of this work. E.F.H. was partially funded by the In-house Laboratory Independent Research (ILIR) program, Program Element 0601153N, managed by the NSWC Carderock Division Director of Research for the ONR. Further, the authors acknowledge the University of Virginia Nanoscale Materials Characterization Facility for the use of the PHI Versaprobe III XPS and SEM, the Eyring Materials Center at Arizona State University supported in part by NNCI-ECCS-1542160 for the WDS work, and the Johns Hopkins University Materials Characterization and Processing Center, in the Whiting School of Engineering for use of Focused Ion Beam and Transmission Electron Microscopy tools. The PHI VersaProbe III system was supported by NSF-MRI Award #162601, MRI Acquisition of an X-ray Photoelectron Spectrometer for Chemical Mapping of Evolving Surfaces: A Regional Instrument for Research and Teaching. Thanks also to Dan Gopman for dicing the wafers at the NIST Center for Nanoscale Science and Technology. This research used the NIST Beamline for Materials Measurement of the National Synchrotron Light Source II, a U.S. Department of Energy (DOE) Office of Science User Facility operated for the DOE Office of Science by Brookhaven National Laboratory under Contract No. DE-SC0012704. We thank Bruce Ravel for assistance in collecting the EXAFS data.

**Author Contributions** DS contributed to Conceptualization, Investigation, Formal analysis, Data Curation, and Writing and Preparation of Original Draft. EFH contributed to Conceptualization, Investigation, Formal analysis, Data Curation, and Writing and Preparation of Original Draft. WHB contributed to Investigation, Formal analysis, and Writing, Reviewing, & Editing of the manuscript. EAA contributed to Formal analysis, Methodology, Software, and Visualization. DLF contributed to Investigation and Data Curation JL contributed to Formal analysis and Supervision. BLDeC contributed to Supervision, Formal analysis, Methodology, Software, and Writing, Reviewing, & Editing of the manuscript. JH-S contributed to Conceptualization, Supervision, Methodology, and Writing, Reviewing, & Editing of the manuscript. KS contributed to Supervision, Methodology, Writing, Reviewing & Editing of the manuscript, and Funding acquisition. HJ contributed to Supervision, Methodology, and Writing, Reviewing, & Editing of the manuscript. JRS contributed to Supervision, Conceptualization, Methodology, Writing—Reviewing, & Editing of the manuscript, and Funding acquisition MLT contributed to Supervision, Conceptualization, Methodology, Writing, Reviewing, & Editing of the manuscript, and Funding acquisition.

**Data Availability** Data can be found here: https://myuva-my.sharepoint.com/:f:/r/personal/ds8vw_virginia_edu/Documents/Old%20combi%20paper/Shared%20Data?csf=1&web=1&e=TgnCpX.

## Declarations

**Competing interests** The authors declare that they have no known competing financial interests or personal relationships that could have appeared to influence the work reported in this paper.

**Open Access** This article is licensed under a Creative Commons Attribution 4.0 International License, which permits use, sharing, adaptation, distribution and reproduction in any medium or format, as long as you give appropriate credit to the original author(s) and the source, provide a link to the Creative Commons licence, and indicate if changes were made. The images or other third party material in this article are included in the article's Creative Commons licence, unless indicated otherwise in a credit line to the material. If material is not included in the article's Creative Commons licence and your intended use is not permitted by statutory regulation or exceeds the permitted use, you will need to obtain permission directly from the copyright holder. To view a copy of this licence, visit http://creativecommons.org/licenses/by/4.0/.

**Publisher's Note** Springer Nature remains neutral with regard to jurisdictional claims in published maps and institutional affiliations.

## Authors and Affiliations

Debashish Sur[1,2] · Emily F. Holcombe[3,4] · William H. Blades[5] · Elaf A. Anber[3] · Daniel L. Foley[3] · Brian L. DeCost[6] · Jing Liu[7] · Jason Hattrick-Simpers[6,8] · Karl Sieradzki[5] · Howie Joress[6] · John R. Scully[1,2] · Mitra L. Taheri[3]

✉ Mitra L. Taheri
mtaheri4@jhu.edu

[1] Department of Materials Science and Engineering, University of Virginia, Charlottesville, VA 22904, USA

[2] Center for Electrochemical Science and Engineering, University of Virginia, Charlottesville, VA 22904, USA

[3] Department of Materials Science and Engineering, Johns Hopkins University, Baltimore, MD 21218, USA

[4] Naval Surface Warfare Center, Carderock, West Bethesda, MD 20817, USA

[5] Ira A. Fulton School of Engineering, Arizona State University, Tempe, AZ 85287, USA

[6] Materials Measurement Science Division, National Institute of Standards and Technology, Gaithersburg, MD 20899, USA

[7] Department of Physics and Astronomy, Manhattan College, Riverdale, New York, NY 10471, USA

[8] Department of Materials Science, University of Toronto, Toronto, ON M5S 3E4, Canada